\documentclass[11pt,a4paper]{article}
\usepackage{jcappub}
\def\vect#1{\mbox{\boldmath $#1$}}

\title{Universal upper limit on inflation energy scale from cosmic
magnetic field}
\author[a,b]{Tomohiro Fujita}
\author[a]{Shinji Mukohyama}
\affiliation[a]{Kavli Institute for the Physics and Mathematics of the
Universe (Kavli IPMU), TODIAS,  the University of Tokyo,\\ 5-1-5
Kashiwanoha, Kashiwa, 277-8583, Japan}
\affiliation[b]{Department of Physics, University of Tokyo,\\ Bunkyo-ku
113-0033, Japan}

\emailAdd{tomohiro.fujita@ipmu.jp}
\emailAdd{shinji.mukohyama@ipmu.jp}

\abstract{
Recently observational lower bounds on the strength of cosmic magnetic
fields were reported, based on $\gamma$-ray flux from distant blazars. 
If inflation is responsible for the generation of such magnetic
fields then the inflation energy scale is bounded from above as 
$\rho_{\rm inf}^{1/4}
<2.5 \times 10^{-7}M_{\rm Pl}\times (B_{\rm obs}/10^{-15}{\rm G})^{-2}$ 
in a wide class of inflationary magnetogenesis models, where 
$B_{\rm obs}$ is the observed strength of cosmic magnetic fields. 
The tensor-to-scalar ratio is correspondingly constrained as 
$r< 10^{-19}\times (B_{\rm obs}/10^{-15}{\rm G})^{-8}$. Therefore, if the 
reported strength $B_{\rm obs}\geq 10^{-15}{\rm G}$ is confirmed and if any
signatures of gravitational waves from inflation are detected in the
near future, then our result indicates some tensions between
inflationary magnetogenesis and observations. 
}

\arxivnumber{1205.5031}

\begin{document}

\maketitle

\section{Introduction}\label{sec:intro}

In 2010, the first detection of cosmic magnetic fields was reported
~\cite{Neronov:1900zz} (see also \cite{Tavecchio:2010mk, Dermer:2010mm,
Huan:2011kp, Dolag:2010ni, Essey:2010nd, Taylor:2011bn, Vovk:2011aa,
Takahashi:2011ac}).  Although High Energy Stereoscopic System (HESS)
$\gamma$-ray telescopes observed TeV scale $\gamma$-rays from several 
blazars, Fermi space telescope did not observe GeV scale $\gamma$-rays
from same blazars. Without cosmic magnetic fields, these two
observations would contradict each other because some parts of TeV scale
$\gamma$-rays traveling through the inter-galactic medium are converted
into GeV scale $\gamma$-rays by electromagnetic cascade reaction; TeV
scale $\gamma$-rays emit electron/positron pairs by scattering with
extragalactic background lights and then created electrons and positrons
emit GeV scale $\gamma$-rays by inverse Compton scattering with CMB
photons until traveling $\mathcal{O}(1)$Mpc
typically~\cite{Huan:2011kp}. In the presence of cosmic magnetic fields,
on the other hand, they can bend the trajectory of charged particles and 
consequently decrease the flux of secondary GeV scale
$\gamma$-rays. 
\footnote{Recently, refs.\cite{Broderick:2011av, Chang:2011bf,
 Pfrommer:2011bg} contested the reliability of this mechanism, pointing
 out the possibility that the low flux of GeV scale $\gamma$-rays may
 not be due to cosmic magnetic field but may be due to a plasma beam
 instability. However, more recently, the Monte Carlo simulation in
 ref.\cite{Miniati:2012ge} showed that the timescale of such 
 instability is much longer than the timescale of electromagnetic
 cascade. Thus, according to them, the plasma beam instability cannot
 explain the lack of GeV scale $\gamma$-rays.} 
From the observational lower limit on the bending angle 
the lower limit on the cosmic magnetic field strength was
obtained. Several works were devoted to this
subject~\cite{Neronov:1900zz,Tavecchio:2010mk, Dermer:2010mm,
Huan:2011kp, Dolag:2010ni, Essey:2010nd, Taylor:2011bn, Vovk:2011aa, 
Takahashi:2011ac}. Some of them took account of possible time variance of 
intrinsic blazar fluxes. The reported lower limit ranges
$\mathcal{O}(10^{-14}$--$10^{-20})$G .

The problem is what the origin of cosmic magnetic fields is. No  
astrophysical process or early universe phenomenology is known to
explain sufficient amount of magnetogenesis~\cite{Kandus:2010nw}. As 
for the inflationary magnetogenesis, many scenarios were proposed,
aiming to explain the origin of magnetic fields of galaxies or galaxy
clusters as well as cosmic magnetic fields~\cite{Turner:1987bw,
Ratra:1991bn, Gasperini:1995dh, Finelli:2000sh, Davis:2000zp,
Giovannini:2000wta, Bassett:2000aw, Bamba:2003av, Bamba:2008hr,
Bamba:2008my, Durrer:2010mq}. The major obstacle in those models is the
so-called ``back reaction problem''~\cite{Bamba:2003av, Kanno:2009ei, 
Demozzi:2009fu}. Generation of magnetic fields during inflation
inevitably increases the energy density of the electromagnetic field. If
it becomes comparable with the energy density of inflaton then the
dynamics of inflation is significantly altered. Consequently, the
inflationary epoch may end or generation of magnetic fields is
drastically suppressed. Demozzi, et al.~\cite{Demozzi:2009fu} pointed
out that in some specific models this problem is crucial and prevents
generation of sufficient magnetic fields.

In the present paper we conduct a model independent analysis of
inflationary magnetogenesis. Specifically, we derive an upper limit on
the inflation energy scale by assuming that all observed cosmic magnetic
fields are generated during inflation. Our constraint depends on neither
details of the model lagrangian, the behavior of photon mode functions
nor the shape of magnetic field spectrum. If the strength of cosmic
magnetic fields is stronger than $10^{-15}$G, the upper limit on the
inflation energy density is $2.5 \times 10^{-7} M_{{\rm Pl}}$. As a consequence,
tensor-to-scalar ratio $r$ is severely constraint as $r<10^{-19}$. Therefore in this case, if all observed cosmic magnetic
fields are generated during inflation, it is almost impossible to detect
gravitational waves from inflation in near future
observations. Conversely, if any signatures of inflationary
gravitational waves are detected, then the cosmic magnetic fields should
have another origin.

The rest of the paper is organized as follows. In section \ref{sec:obs},
we introduce the observational lower limit on the strength of the
magnetic field. In section \ref{sec:asmp}, we discuss the assumptions
which are needed to derive the main result. In section \ref{sec:lim}, the
derivation of the upper limit on the inflation energy scale is
presented. In section \ref{sec:int}, we investigate the validity of the third
assumption and explore the possibility to evade the constraint. Section
\ref{sec:con} is devoted to a summary of this paper. In appendix, we
derive the observational lower bound of cosmic magnetic fields in terms
of the magnetic power spectrum.

\section{Observational Constraint on Magnetic Power Spectrum} \label{sec:obs}

From the observations of blazars, current strength of cosmic magnetic
fields is constrained ~\cite{Neronov:1900zz,Tavecchio:2010mk,
Dermer:2010mm, Huan:2011kp, Dolag:2010ni, Essey:2010nd, Taylor:2011bn,
Vovk:2011aa, Takahashi:2011ac}. While the constraints in those
literatures are given in terms of the correlation length of magnetic fields, it is 
straightforward to rewrite it for the power spectrum of magnetic fields as 
\begin{equation}
 B_{\rm eff}^2(\eta_{\rm now})
  \equiv \int^{k_{\rm diff}}_0 \frac{{\rm d} k}{k} 
  F(kL)\, \mathcal{P}_{B}(\eta_{\rm now},k)
 \ \geq \ 
 B_{\rm obs}^{2},
  \label{eq:observational limit}
\end{equation}
where $\mathcal{P}_{B}(\eta_{\rm now},k)$ is the power spectrum of
the magnetic field at the present, 
\begin{equation}
 F(z) \equiv \frac{3}{2}z^{-2}
  \left[ \cos (z) - \frac{\sin (z)}{z} + z {\rm Si}(z) \right],
\end{equation}
and Si$(z)$ denotes the sine integral function. (See appendix for
derivation. For the introduction of $k_{\rm diff}$, see the explanation
below.) Here, $\eta$ is the conformal time, the subscript "now" denotes
the present value, $L\equiv 1{\rm Mpc}$ stands for the  characteristic 
length scale for energy losses of charged particles due to inverse
Compton scattering ~\cite{Huan:2011kp}, $B_{\rm obs}$ is the
observational lower limit on the strength of cosmic magnetic fields, and
$k_{\rm diff} \simeq (1{\rm AU})^{-1}$ is the wave number corresponding 
to the present cosmic diffusion length, i.e. the minimal size of a magnetic
configuration which can survive diffusion during the Universe's
lifetime~\cite{Grasso:2000wj}. Note that the effect of the cosmic 
diffusion can be expressed as the cutoff in the power spectrum as
\begin{equation}
 \mathcal{P}_{B}(\eta_{\rm now},k) \simeq 0 \quad 
  (\mbox{for } k>k_{\rm diff}). \label{eqn:cutoff-diff}
\end{equation}
in cases where magnetic fields have their origin in the primordial universe.
In the formula (\ref{eq:observational limit}), the cutoff at 
$k_{\rm diff}$ is made more explicit as the domain of integration.

For $z \geq 0$, $F(z)$ satisfies 
\begin{equation}
 0< F(z) \leq 1, \quad 
  0 \leq zF(z) \leq \alpha, \quad 
  \alpha\equiv {\rm Max} [zF(z)] \simeq 2.48.
  \label{eqn:F_inequalities}
\end{equation}
Although $B_{\rm obs}$ still has a few orders of uncertainty, in the
present paper we adopt the value reported by ref.\cite{Taylor:2011bn}
in which the latest data were analyzed. According to
ref.\cite{Taylor:2011bn}, $B_{\rm obs} \simeq 10^{-15}$G unless the
time variance of intrinsic blazar fluxes is significant.

The derivation of the formula (\ref{eq:observational limit}) is given in
appendix. Here, instead of showing the detailed derivation, we provide
intuitive understanding of it. For this purpose, let us replace $F(kL)$
by its asymptotic forms,
\begin{equation}
  F(z) \sim \left\{
\begin{array}{lc}
 1 + \mathcal{O} (z^2) & \quad (z \ll 1) \\
 \frac{3\pi}{4z} + \mathcal{O} (z^{-2}) & \quad (z \gg 1)
  \end{array} \right. ,
\end{equation}
and drop $\mathcal{O} (1)$ numerical factors to obtain the approximate
formula as
\begin{equation}
 B_{\rm eff}^{2}(\eta_{\rm now}) 
  \sim
  \int_0^{1/L} \frac{{\rm d} k}{k} \biggl[\mathcal{P}_{B}
  (\eta_{\rm now},k) \biggr] 
  +
  \int^{k_{\rm diff}}_{1/L} \frac{{\rm d} k}{k} 
  \left[ \frac{1}{kL} \mathcal{P}_{B} (\eta_{\rm now},k)
  \right]. \label{eq:Beff_approx}
\end{equation}
Let us now think of a Fourier mode of the magnetic field. For $kL\ll 1$,
the corresponding magnetic field can be treated as a homogeneous field,
as far as the particle's trajectory (with the total length $L$) is
concerned. Thus modes with $kL\ll 1$ contribute to the bending angle as
if they are homogeneous fields. This explains the first term in the
right hand side of (\ref{eq:Beff_approx}). On the other hand, for 
$kL\gg 1$, the direction of the corresponding magnetic field randomly
changes $N\sim kL$ times while the charged particle travels the total
length $L$. If we were interested in the trajectory of the charged
particle within one of short segments of the length $\sim k^{-1}$ then
the magnetic field could be treated as a homogeneous field. Actually, we
are interested in the total bending angle due to $N$ segments. Because
of the randomness of the direction, the total bending angle from $N$ 
segments adds up to only $\sqrt{N}$ times the contribution from each
segment. Therefore the contribution of modes with $kL\gg 1$ to the
variance of the bending angle should acquire the weight of order 
$1/N\sim 1/(kL)$. This explains the second term in the right hand side
of (\ref{eq:Beff_approx}).

\section{Four Assumptions}\label{sec:asmp}

To derive the upper limit on the inflation energy scale, we need four
assumptions.

\subsection{Assumption 1: the form of kinetic term}

First, we assume that the kinetic term of the photon field $A_\mu$ is of
the form 
\begin{equation}
 {\mathcal{L}}_{\rm kin} = 
  -\frac{1}{4} I^2(\eta) F_{\mu\nu}F^{\mu\nu}, 
  \label{eq:first assumption}
\end{equation}
where it is understood that the time-dependence of $I(\eta)$ is due to
its dependence on homogeneous, time-dependent fields present in the
theory. Thus, ${\mathcal{L}}_{\rm kin}$ includes various interactions 
between the photon field and other fields~\cite{Ratra:1991bn,
Gasperini:1995dh, Bamba:2003av, Demozzi:2009fu}. This form of coupling
does not have to break either gauge or local Lorentz symmetry. In
general the photon field can have additional interactions 
${\mathcal{L}}_{\rm int}$:
\begin{equation}
 {\mathcal{L}}_A = 
  {\mathcal{L}}_{\rm kin} + {\mathcal{L}}_{\rm int}. 
\end{equation}
However we let ${\mathcal{L}}_{\rm int}$ 
unspecified. Even so, under the four assumptions introduced in this
section, we can derive the upper limit on the inflation energy scale in 
a model independent way. Note that when $I=1$ and
${\mathcal{L}}_{\rm int}=0$, the usual Maxwell theory is restored.

This assumption on the form of the kinetic term is necessary to 
quantize the photon field and to define the kinetic energy density. The
photon field $A_\mu$ can be separated into scalar and vector modes as 
\begin{equation}
A_{\mu}(\eta, \mathbf{x})
 =\left( A_0, V_i+\partial_i S \right),~~~~~~~\partial_i V_i=0,
\label{eq:def of vector mode}
\end{equation}
where $A_0$ and $S$ are the scalar modes and $V_i$ are the vector
modes. Let us quantize the vector modes. After Fourier transformation
with respect to the spatial coordinates, expansion by polarization
vectors and mode expansion, we impose the standard commutation relation
on the creation and annihilation operators. 
\begin{eqnarray}
 V_i(\eta, \mathbf{x})
 & = & 
 \sum_{p=1}^{2} \int \frac{{\rm d}^3 k}{(2\pi)^3} 
 e^{i \mathbf{k \cdot x}} \epsilon_{i}^{(p)}(\hat{\mathbf{k}}) 
 \left[ a_{\mathbf{k}}^{(p)} u_{k}^{(p)}(\eta) 
  + a_{\mathbf{-k}}^{\dag (p)} u_{k}^{(p)*}(\eta) \right]\\
 \left[ a_{\mathbf{k}}^{(p)}, 
  a_{\mathbf{k}^{\prime}}^{\dag (p^{\prime})} \right]
  & = & 
  (2\pi)^3 {\delta}^{(3)}(\mathbf{k-k^{\prime}}) {\delta}^{p p^{\prime}},
\end{eqnarray}
where $u_k(\eta)$ is the mode function of the photon vector mode, $p$
($=1,2$) is the polarization label and 
${\epsilon}_{i}^{(p)}(\hat{\mathbf{k}})$ is the polarization vector
satisfying 
\begin{equation}
k_i \epsilon_{i}^{(p)}(\hat{\mathbf{k}})=0~~~(p=1,2) ,~~~~~~
\sum_{p=1}^{2} {\epsilon}_{i}^{(p)}(\hat{\mathbf{k}}) {\epsilon}_{j}^{(p)}(-\hat{\mathbf{k}})=\delta_{ij} - \frac{k_{i}k_{j}}{k^2}.
\end{equation}
Then the canonical commutation relation for $V_i$ requires the
normalization condition of mode function $u_{k}(\eta)$ as 
\begin{equation}
 I^{2}
  \left( u_{k}^{(p)}\, \partial_{\eta}u_{k}^{(p)*}
   - u_{k}^{(p)*}\, \partial_{\eta}u_{k}^{(p)} \right )
   = i~~~~~(p=1, 2). 
\label{eq:normalization condition}
\end{equation}

Now let us define the contribution of modes with $k<k_{\rm diff}$,
i.e. those whose comoving length scale are longer than the cosmic
diffusion length, to the kinetic energy density of the electromagnetic
field as~\footnote{
It is understood that the domain of integration over $k$ in
eq.(\ref{eq:kin energy}) is the same as in 
eq.(\ref{eq:observational limit}). As a result, 
$\rho_{\rm kin}(k_{\rm diff},\eta)$ is not the sum of all modes that
exist during inflation but the sum of modes whose scales are relevant to
observed cosmic magnetic fields.} 
\begin{equation}
 \rho_{\rm kin} (k_{\rm diff}, \eta) 
  = \frac{I^2}{2} \int^{k_{\rm diff}}_0
\frac{{\rm d} k}{k}
  \left[{\mathcal{P}}_E (\eta, k)
   +{\mathcal{P}}_B (\eta, k) \right],
 \label{eq:kin energy}
\end{equation}
where we have defined power spectra of electric and magnetic fields as 
\begin{equation}
{\mathcal{P}}_E (\eta, k)
 =\frac{k^3 |u_{k}^{\prime}(\eta)|^2}{\pi^2 a^4(\eta)},\quad
 {\mathcal{P}}_B (\eta, k)
 =\frac{k^5 |u_{k}(\eta)|^2}{\pi^2 a^4(\eta)}. 
\label{eq:def of power spectrum}
\end{equation}

\subsection{Assumption 2: Avoidance of strong coupling}

The second assumption is that 
\begin{equation}
 I(\eta) \ge 1 \quad \mbox{for} \ \eta_{\rm diff} \leq \eta,
  \label{eq:second assumption}
\end{equation}
where $\eta_{\rm diff}$ is the conformal time when the comoving cosimic diffusion
length exits the horizon, thus defined as $k_{\rm diff} \eta_{\rm diff} = -1$.

This assumption essentially states that the effective coupling constants
of the photon field to other fields should be always smaller than
present values. For example, let us consider the interaction between the 
photon and a charged fermion as  
\begin{equation}
 {\mathcal{L}}_{\rm int} \ni 
  -e \bar{\psi} \gamma^{\mu} \psi A_\mu.
\end{equation}
In order to evaluate the effective coupling constant, we should
canonically normalize the fields. Let us suppose that the fermion $\psi$
is already canonically normalized. The canonically normalized photon
field is $A^c_{\mu} \equiv I A_{\mu}$. Then the interaction term is
rewritten as 
\begin{equation}
 {\mathcal{L}}_{\rm int} \ni 
  -\frac{e}{I} 
  \bar{\psi} \gamma^{\mu} \psi A^c_{\mu}. 
\end{equation}
It is now clear that $e/I$ is the effective coupling constant. Therefore
if $I \ll 1$, the effective coupling constant becomes large and the tree
level analysis would be invalidated. In order to justify the tree level
analysis, we need to assume that $I$ is bounded from below by a positive
constant $I_0$. For simplicity we set $I_0$ to be the present value of
$I$, i.e. $I_0=1$.

\subsection{Assumption 3: Small back reaction}

The third assumption is that the kinetic energy density eq.(\ref{eq:kin energy})
is smaller than that of inflaton
\footnote{We are interested in modes with $k<k_{\rm diff}$ only. One can show
that these modes do not significantly contribute to the
vacuum enegy part of $\rho_{\rm kin}$.
Validity of the effective field theory requires that the
contribution of each mode $\omega_{k}$ to the vacuum energy must be
smaller than the Planck mass scale $M_{\rm Pl}$. Hence, 
$|\rho_{\rm vac}(k_{\rm diff}, \eta)| < 
({\rm Max}_{k\leq k_{\rm diff}}|\omega_k|) (k_{\rm diff}/a)^3 < 
M_{\rm Pl} H^3 \ll \rho_{\rm inf}$ for 
$\eta_{\rm diff} \leq \eta\leq \eta_f$. 
Therefore distinction between renormalized and unrenormalized
expressions is irrelevant for (\ref{eq:kin energy}). One can show that
such distinction is unimportant also for (\ref{eq:observational limit}).
}, 
\begin{equation}
 \rho_{\rm kin} (k_{\rm diff}, \eta)  <  
  \rho_{\rm inf} \quad \mbox{for}\ 
  \eta_{\rm diff}\leq \eta\leq \eta_f, 
\label{eq:third assumption}
\end{equation}
where $\eta_f$ is the conformal time at
the end of inflation and hereafter we ignore the time-dependence of the
inflaton energy density $\rho_{\rm inf}$.

This assumption is closely related to the condition for avoidance of the
back reaction problem
\begin{equation}
|\rho_{\rm kin}(\eta) + \rho_{\rm int}(\eta)| 
 < \rho_{\rm inf} \quad \mbox{for}\ 
  \eta_{\rm diff}\leq \eta\leq \eta_f.  
\label{eq:continuance condition}
\end{equation}
Note that eq.(\ref{eq:third assumption}) and 
eq.(\ref{eq:continuance condition}) are different. In general, the total
energy density of the photon field includes not only the kinetic energy
density $\rho_{\rm kin}$ but also the interaction energy density 
$\rho_{\rm int}$ due to the additional interaction terms
${\mathcal{L}}_{\rm int}$. Also, $\rho_{\rm kin}(\eta)$ in
(\ref{eq:continuance condition}) should be understood as 
$\rho_{\rm kin}(\infty,\eta)$ and thus is in general larger than
$\rho_{\rm kin}(k_{\rm diff},\eta)$ in (\ref{eq:third assumption}). If
the interaction energy density is non-negative ($\rho_{\rm int}\geq 0$)
then eq.(\ref{eq:continuance condition}) requires eq.(\ref{eq:third
assumption}). Even if the interaction energy is 
negative ($\rho_{\rm int}< 0$), unless the two contributions 
$\rho_{\rm kin}$ and $\rho_{\rm int}$ cancel each other with a
sufficiently good precision, eq.(\ref{eq:continuance condition})
generically requires eq.(\ref{eq:third assumption}). Therefore the third
assumption eq.(\ref{eq:third assumption}) is mandatory unless negative  
$\rho_{\rm int}$ precisely cancels out positive $\rho_{\rm kin}$.

In section \ref{sec:int}, we confirm the necessity of the third
assumption in the case of gauge and local Lorentz invariant quadratic 
interactions and explore the possibility of the precise cancellation 
between $\rho_{\rm kin}$ and $\rho_{\rm int}$.

\subsection{Assumption 4: Magnetogenesis during inflation}

The fourth assumption is that all observed magnetic fields are generated
during inflation. In particular, the conformal symmetry of the photon
field action is broken appreciably only in the inflationary era. Since
the electric conductivity of the universe increases after the end 
of inflation~\cite{Turner:1987bw} (see also (\ref{eqn:cutoff-diff})), we
have 
\begin{equation}
 B_{\rm eff}^2(\eta_{\rm now}) 
  \leq a_f^4 B_{{\rm eff}}^2 (\eta_{f}), 
\label{eq:fourth assumption one}
\end{equation}
where we have set $a(\eta_{\rm now})=1$.

By using eq.(\ref{eq:second assumption}), (\ref{eq:third assumption})
and the fact that $B_{\rm eff}$ is smaller than the usual definition of
magnetic field strength ($0<F(kL)\leq 1$), we obtain 
\begin{equation}
B_{\rm eff}^2(\eta_{\rm diff}) < 2 \rho_{\rm inf}.
\label{eq:beginning relation}
\end{equation}
Assuming the instantaneous reheating, we find the scale factor at the
end of inflation is given by 
 \begin{equation}
  a_f^4 = \frac{\rho_{\gamma}}{\rho_{\rm inf}}
 \label{eq:af}
\end{equation}
where 
$\rho_{\gamma} \simeq 5.7 \times 10^{-125} M_{\rm Pl}^{4}
\simeq 5.2 \times 10^{-12} {\rm G}^2$ 
is the present energy density of radiation. 
Eq.(\ref{eq:fourth assumption one}), 
(\ref{eq:beginning relation}), (\ref{eq:af}) and (\ref{eq:observational limit})
lead to the following inequality. 
\begin{equation}
 \frac{a_{\rm diff}^4 B_{\rm eff}^2(\eta_{\rm diff})}{a_f^4 B_{\rm eff}^2(\eta_f)} 
  <
  10^{-42} \times \exp[-4(\Delta N-35)] 
  \left( \frac{B_{\rm obs}}{10^{-15}G} \right)^{-2},
\label{eq:negligible}
\end{equation}
where $\Delta N \equiv \ln(a_f/a_{\rm diff})$. This inequality implies that
\begin{equation}
 a_{\rm diff}^4 B_{\rm eff}^2(\eta_{\rm diff}) \ll a_f^4 B_{\rm eff}^2(\eta_f),
  \label{eq:fourth assumption two}
\end{equation}
and thus states that the magnetic fields have to be significantly
amplified during inflation to explain the observational lower limit
eq.(\ref{eq:observational limit}).

\section{Upper Limit on Inflation Energy Scale}\label{sec:lim}

With the four assumptions stated in the previous section, we are now
ready to derive the upper limit on the inflation energy scale. The
derivation is independent of details of inflationary magnetogenesis
models, the behavior of photon mode functions or the spectrum of the
electromagnetic fields.

Independently from the specific functional form of $u_k (\eta)$, it can
be shown that 
\begin{eqnarray}
 |u_k(\eta_{f})|^2-|u_k(\eta_{\rm diff})|^2 
  &=& 
  \int^{\eta_{f}}_{\eta_{\rm diff}} {\rm d}\eta ~ 
  2 |u_k (\eta)|\, |u_{k} (\eta)|^{\prime}
  \nonumber \\
 &\le&
  \int^{\eta_{f}}_{\eta_{\rm diff}}  \frac{{\rm d}\eta}{k} ~ 
  2 k|u_k (\eta)|\, |u_{k}^{\prime} (\eta)| 
  \nonumber\\
 &\le&
  \int^{\eta_{f}}_{\eta_{\rm diff}} \frac{{\rm d}\eta}{k} 
  \biggl( k^2 | u_k (\eta) |^2 + |u_k^{\prime}(\eta)|^2 \biggr),
  \label{eq:soukasoujou}
\end{eqnarray}
where we have used the general inequalities $|z(\eta)|^\prime \le |z^\prime (\eta)|$ for
a complex function $z(\eta)$ and $2xy \le x^2+y^2$ for real numbers $x$ 
and $y$. Multiplying the both ends of eq.(\ref{eq:soukasoujou}) by 
$F(kL)k^4/\pi^2$ and integrating it over $k$ from $0$ to $k_{\rm diff}$,
we obtain 
\begin{equation}
 a_f^4 B_{\rm eff}^2 (\eta_f) - a_{\rm diff}^4 B_{\rm eff}^2 (\eta_{\rm diff})
  < 
  \frac{\alpha}{L}\int^{\eta_{f}}_{\eta_{\rm diff}}{\rm d}\eta ~ 
  a^4 (\eta) \int_0^{k_{\rm diff}}\frac{{\rm d} k}{k} 
  \left[\mathcal{P}_E (\eta,k) + \mathcal{P}_B (\eta,k)\right],
\end{equation}
where we have used the second inequality listed in
(\ref{eqn:F_inequalities}). Using the second, third and fourth
assumptions as well as eq.(\ref{eq:fourth assumption two}), we obtain  
\begin{equation}
 B_{\rm eff}^2(\eta_{\rm now})
  < 
  \frac{2\alpha}{L} \rho_{\rm inf}
  \int^{\eta_f}_{\eta_{\rm diff}} d\eta\, a^4(\eta)
  \simeq
  \frac{2\alpha}{3H_{\rm inf}L} a_{f}^{3} \rho_{\rm inf}
  \label{eq:semifinal inequality}
\end{equation}
where $H_{\rm inf}$ ($\simeq$ const.) is the Hubble expansion rate
during inflation.

Note that $1/H_{\rm inf}$ and $a_f^3$, which appear in the r.h.s. of
(\ref{eq:semifinal inequality}), are decreasing functions of the
inflation scale. Indeed, by substituting (\ref{eq:af}) for $a_f$ and
using the Friedmann equation for $H_{\rm inf}$, we can see that the
r.h.s. of (\ref{eq:semifinal inequality}) is a decreasing function of
the inflation scale. Hence, substituting eq.(\ref{eq:observational
limit}) and eq.(\ref{eq:af}) into eq.(\ref{eq:semifinal inequality}), we
finally obtain the upper limit on the inflation energy scale, 
\begin{equation}
\rho^{1/4}_{{\rm inf} }
 <
 \frac{2\alpha}{\sqrt{3}L}
 \rho^{3/4}_{\gamma} M_{{\rm Pl}}B_{\rm obs}^{-2}
 \approx
 2.5 \times 10^{-7} M_{{\rm Pl}} \times
 \left( \frac{B_{\rm obs}}{10^{-15}G} \right)^{-2}.
\label{eq:main result}
\end{equation}
Note this upper limit can become even stronger if details of
reheating is taken into consideration instead of
eq.(\ref{eq:af})~\cite{Demozzi:2012wh}. Provided that the dominant energy density behaves like
matter $(\propto a^{-3})$ during reheating, the right-hand side of
eq.(\ref{eq:main result}) is multiplied by an additional factor
$(\rho_{\rm reh}/\rho_{\rm inf})^{1/4}<1$, where 
$\rho_{\rm reh}$ is the energy density at the end of reheating era.

Eq.(\ref{eq:main result}) can be converted into the upper bound on the
tensor-to-scalar ratio $r$ under the slow-roll approximation, 
\begin{equation}
 r < 10^{-19} \times 
  \left( \frac{B_{\rm obs}}{10^{-15}G} \right)^{-8}.
  \label{eq:r limit}
\end{equation}
Therefore, if all observed cosmic magnetic fields are generated during 
inflation, it is extremely difficult to detect any signatures of
primordial gravitational waves, for example direct detections or CMB
B mode polarization. Conversely, if some observations reveal that $r$ is
larger than the upper bound (\ref{eq:r limit}), it implies that
inflation cannot explain the origin of cosmic magnetic fields
under the four assumptions.

Now let us discuss the intuitive understanding of the reason why we
obtain the upper limit on the inflation energy scale. 
Roughly speaking, $a^4 \mathcal{P}_{B}$ has to increase significantly
during inflation for inflationary magnetogenesis 
(see eq.(\ref{eq:fourth assumption two})).
It is easy to show in the same way as eq.(\ref{eq:soukasoujou}) that
\begin{equation}
\frac{1}{k} \frac{d}{d\eta} \left( a^4(\eta) \mathcal{P}_{B} (\eta,k) \right)
\le
a^4(\eta) \left( \mathcal{P}_{E}(\eta,k) + \mathcal{P}_{B}(\eta,k) \right).
\label{eq:intuitive understanding}
\end{equation}
From the second and third assumption we know right-hand side of 
eq.(\ref{eq:intuitive understanding}) integrated by ln$k$ should be
smaller than $a^4 \rho_{\rm inf}$. Thus essentially, time variation
 of $a^4 \mathcal{P}_{B}$ is bounded from above by $a^4 \rho_{\rm inf}$.
Then we can rewrite $a^4 \rho_{\rm inf}$ by using eq.(\ref{eq:af}) as
\begin{equation}
a^4 (\eta) \rho_{\rm inf} = \rho_{\gamma} e^{-4N},
\quad
N \equiv \sqrt{\frac{\rho_{\rm inf}}{3 M^2_{\rm Pl}}} (t_{\rm f}-t).
\end{equation}
where $N$ is e-folding number, $t$ is cosmic time and $t_{\rm f}$
denotes the end of inlation. Therefore $a^4 \rho_{\rm inf}$ is actually
\emph{decreasing function} of $\rho_{\rm inf}$ during inflation.
Since lower $\rho_{\rm inf}$ is favored to relax the upper bound on
time variation of $a^4 \mathcal{P}_{B}$, we obtain the upper
bound on inflation energy scale.

\section{Additional Interaction Terms}\label{sec:int}

The action for the photon field consists of not only the kinetic term
${\mathcal{L}}_{\rm kin}$ but also the additional interaction terms
${\mathcal{L}}_{\rm int}$. As already mentioned after 
(\ref{eq:continuance condition}), the third assumption eq.(\ref{eq:third
assumption}) is mandatory unless negative $\rho_{\rm int}$ precisely
cancels out positive $\rho_{\rm kin}$. Therefore whether such a precise
cancellation is possible is a significant question. The answer we shall
draw in the following discussion is that it is rather difficult to
achieve such a cancellation.

Here, it is perhaps worthwhile stressing that, as long as the four
assumptions (including the third one) are satisfied, our main result 
eq.(\ref{eq:main result}) holds even if $\rho_{\rm int}$ and 
$\rho_{\rm kin}$ precisely cancel out.

\subsection{Gauge and Lorentz invariant quadratic term}

In the quadratic level, the most general renormalizable interaction
term which preserves gauge and local Lorentz symmetry is given by 
\begin{equation}
 \mathcal{L}_{\rm int}
  =
  \frac{1}{8} f(\eta) \epsilon^{\mu\nu\rho\sigma}
  F_{\mu\nu} F_{\rho \sigma}
  +
  \frac{1}{2} m^{2} (\eta) A_{\mu} A^{\mu},
\end{equation}
where $\epsilon^{\mu\nu\rho\sigma}$ is the totally anti-symmetric tensor
with $\epsilon^{0123}=1/\sqrt{-g}$, $f(\eta)$ is a function of
homogeneous scalars. The first term is called axial coupling term. The
second term is the effective mass term of the photon induced by
expectation values of charged scalars. It stems from the kinetic term of
the charged scalars, and the positivity of the time kinetic term implies
the positivity of the mass squared $m^2$. This term spontaneously breaks
the $U(1)$ gauge symmetry, and the longitudinal mode of photon field
becomes a physical degree of freedom.

Actually, the axial coupling term does not contribute to the energy
density of the photon field. Since the axial coupling term does not
depend on the metric, its contribution to the energy momentum tensor
is exactly zero. The effective mass term does contribute to
$\rho_{\rm int}$ but the contribution is always positive because of the
positivity of the mass squared. Therefore the cancellation between
$\rho_{\rm int}$ and $\rho_{\rm kin}$ cannot occur.

\subsection{Model with negative interaction energy}

There is an existing model which gives a \emph{negative} energy
contribution from an additional interaction term. Turner and
Widrow~\cite{Turner:1987bw} proposed a model with non-minimal coupling, 
$\mathcal{L}_{\rm int} \propto R A_{\mu} A^{\mu}$, where 
$R$ is the Ricchi scalar. This coupling can become an effective mass term
of photon with negative mass squared. However this model has three
critical problems. First, the longitudinal mode of photon becomes ghost 
~\cite{Himmetoglu:2008zp, Himmetoglu:2009qi, Demozzi:2009fu}. 
Second, the negative energy contribution from 
$\mathcal{L}_{\rm int}$ exceeds $\rho_{\rm inf}$ and the back reaction 
spoils inflation when we require generated magnetic field is sufficient
~\cite{Demozzi:2009fu}. Third, this coupling explicitly breaks the
gauge symmetry.

\subsection{Energy conserving term}

From purely phenomenological viewpoints, let us investigate the
additional interaction term of the form 
\begin{equation}
\sqrt{-g} \mathcal{L}_{\rm int} = 
 \frac{1}{2} a^2 J^2(\eta) V_{i}^{2}\qquad
 (J^2>0)
\label{eq:test int term}
\end{equation}
where $J(\eta)$ is a function of homogeneous scalar fields and $V_i$ is
the photon vector mode defined in eq.(\ref{eq:def of vector mode}). 
This term is effective mass term of photon vector mode with 
\emph{negative} mass squared. Note that this term has neither gauge
invariance nor Lorentz invariance. It does not yield ghost field because
it contains only vector modes by breaking Lorentz symmetry and we still
assume that the kinetic term of photon is given by eq.(\ref{eq:first
assumption}). Although it may be hard to embed such a term in a viable
elementary particle theory, it is worth investigating it since we can
find an interesting way to realize the cancellation between 
$\rho_{\rm int}$ and $\rho_{\rm kin}$.

From eq.(\ref{eq:test int term}), the equation of motion is given by
\begin{equation}
u_{k}^{\prime\prime} + \left( k^2-a^2 J^2 \right) u_k =0.
\label{eq:EoM}
\end{equation}
Here we have assumed $I(\eta) = 1$ for simplicity, since otherwise the
weak coupling effect due to $I \gg 1$ would make the interaction term
irrelevant. At the same time we require the cancellation between
$\rho_{\rm int}$ and $\rho_{\rm kin}$ for each mode, 
\begin{equation}
|u_{k}^{\prime}|^2 + (k^2-a^2 J^2) |u_k|^2 =0.
\label{eq:cancel condition}
\end{equation}
It is easy to show that eq.(\ref{eq:EoM}), eq.(\ref{eq:cancel
condition}) and eq.(\ref{eq:normalization condition}) imply that 
\begin{equation}
 a^2 J^2(\eta) = {\rm const.}
\end{equation}
In other words, the coefficient of the quadratic term (\ref{eq:test int
term}) should be constant.

The reason why only the interaction term with constant coefficient leads
the cancellation is simple. It is the energy conservation. If there is
no explicit dependence on time in the action (for example, if the
time-dependence due to the scale factor $a(\eta)$ is canceled by
time-evolving scalars), then the energy of the system is conserved by
virtue of Noether's theorem. In the case of eq.(\ref{eq:test int term}),
if $J(\eta)$ cancels the time dependence of $a(\eta)$, the photon energy
(with respect to the conformal time $\eta$) is conserved. Note that the
kinetic term of the photon field is originally free from
$a(\eta)$. Therefore the energy density of photon does not increase even
if the electromagnetic field strength increases. It is notable that, for
this mechanism to work, the dynamics of the scalar fields included in
$J(\eta)$ has to restore the time translation symmetry accidentally.

The above analysis implies that the magnetogenesis from inflation whose
energy is larger than the constraint of eq.(\ref{eq:main result}) may
not be impossible in principle. However, in practice it is not easy to
realize a model which exploits the energy conserving mechanism because
the accidental symmetry restoration by the scalar field dynamics can be
easily spoiled by various effects such as the back reaction of the photon
field. Therefore it is fair to say that all the four assumptions
(including the third one) are likely to be mandatory in a rather broad 
class of models and the derived upper limit on the inflation energy scale is
considerably general.

\section{Conclusion}\label{sec:con}

In this paper we have derived a universal upper limit on the inflation
energy scale under the following four assumptions. (i) The kinetic term
of the photon field is of the canonical form up to a time-dependent
overall factor. (ii) The effective coupling constants do not exceed
present values and thus do not exhibit strong coupling. (iii) The
kinetic energy of the photon field is always lower than the inflaton
energy density during inflation. (iv) All observed cosmic magnetic
fields are generated during inflation.

The derived constraint is eq.(\ref{eq:main result}), 
$\rho^{1/4}_{\rm inf} < 
2.5 \times 10^{-7} M_{{\rm Pl}} \times (B_{\rm obs}/10^{-15}G)^{-2}$. 
As a consequence, the tensor-to-scalar ratio $r$ is bounded from above
as eq.(\ref{eq:r limit}), 
$r < 10^{-19} \times (B_{\rm obs}/10^{-15}G)^{-8}$. We hardly expect
that inflation is the origin of both cosmic magnetic fields and
detectable gravitational waves if $B_{\rm obs}> 10^{-15}G$. Therefore
the future detection of signatures of inflationary gravitational waves,
if any, would imply tension between inflationary magnetogenesis and
observations.

Although our constraint is valid in fairly broad class of inflationary
magnetogenesis scenarios, we have investigated the possibility to evade
it. In order to evade the constraint, at least one of the assumptions
should be violated. The third assumption can be violated only if the
energy density due to additional interaction terms and the kinetic
energy density precisely cancel out. We have considered a possible
mechanism which exploits a energy conservation law to realize the
cancellation. However, it seems a challenge to build a realistic model
equipped with such a mechanism.

So far, we have not succeeded to find a concrete model that produces
relevant magnetic fields during inflation under the four assumptions we
have made. However, this does not necessarily imply non-existence of
such models. In this respect, it is probably worthwhile pointing out
that our four assumptions do not exclude higher dimensional and/or
nonlinear interaction terms. Our discussion can be applied also to
models involving decays of other fields to the electromagnetic
field. Those kinds of models have not been investigated in details
and thus there is a large class of unexplored models. We 
feel that further discussion is beyond the scope of the present paper.

Nonetheless, the result of the present paper is expected to be useful,
providing a new judgment condition. Namely, if tensor-to-scalar ratio
is detected in the future, any possibilities of magnetogenesis model
within our assumptions will be excluded. Alternatively, if one can
derive a lower limit on inflation energy scale by different arguments
for a class of models then the upper limit obtained in this paper can be
used to rule out the class of models. In this sense the upper bound we
have found may be considered as an obstacle to inflationary
magnetogenesis as well as an important guideline for model building.

\acknowledgments

We would like to thank Toshiya Kashiwagi and Takayuki Koike for useful
discussions. This work was supported by the World Premier International
Research Center Initiative (WPI Initiative), MEXT, Japan. 
T.F acknowledges the support by Grant-in-Aid for JSPS Fellows.
S.M. also acknowledges the support by Grant-in-Aid for Scientific Research
24540256 and 21111006, and by Japan-Russia Research Cooperative Program.

\appendix

\section{Derivation of the Constraint on Magnetic Power Spectrum}

In this appendix, we derive eq.(\ref{eq:observational limit}). By Fermi
and HESS observations, there is a lower limit on the bending angle of
GeV scale cascade electrons and positrons in the inter-galactic
medium. However, in the literatures the constraint on the cosmic
magnetic field is given only in terms of the correlation length of magnetic
fields ~\cite{Neronov:1900zz,Tavecchio:2010mk, Dermer:2010mm,
Huan:2011kp, Dolag:2010ni, Essey:2010nd, Taylor:2011bn, Vovk:2011aa, 
Takahashi:2011ac} while theorists need the constraint in terms of the
magnetic power spectrum. In this appendix we shall generalize the
constraint on the cosmic magnetic field to more general spectra. Such a
generalization makes the connection between the cosmic magnetic power
spectrum ${\mathcal{P}}_B$ and the bending angle $\theta$. For
simplicity, we neglect effects of special relativity in this appendix. 

Provided that a charged particle travels distance $L$ in the background
of a weak magnetic field $\vect{B}(\vect{r})$ from $t_1$ till
$t_2$. Then the bending angle is given by 
\begin{equation}
\vect{\theta} \simeq \frac{\vect{v}(t_1)-\vect{v}(t_2)}{v},
\end{equation}
where $\vect{v}(t)$ is the velocity vector of the particle. Note the
absolute value of the velocity vector is constant. By using the equation
of motion with Lorentz force,  the difference of the velocity vectors is
written as 
\begin{equation}
\vect{v}(t_2)-\vect{v}(t_1)
=
\int^{t_2}_{t_1} {\rm d}\tilde{t} ~ {\dot{\vect{v}}} (\tilde{t})
=
\frac{e}{m} \int^{t_2}_{t_1} {\rm d}\tilde{t} ~ \vect{v}(\tilde{t}) \times \vect{B} (\tilde{t})
=
\frac{e}{m} \int^{L}_{0} {\rm d} \vect{x} \times \vect{B}(\vect{x}),
\end{equation}
where $e$ and $m$ are the charge and the mass of the particle,
respectively, $\vect{x}(t)$ denotes the orbit of the particle and its
initial value is set to $\vect{x}(t_1)=0$. Then, we assume $\theta$ is
so small that the orbit can be approximated as a straight line, 
$\vect{x}(t) \simeq x_{1}(t) \hat{\vect{e}}_1$ where 
$\hat{\vect{e}}_1$ is the unit vector in the direction of the axis
$1$. By Fourier transforming $\vect{B}(\vect{x})$, we can perform the
line integral 
\begin{equation}
 \int^{L}_{0} {\rm d} x_1 ~ \hat{\vect{e}}_1 \times \vect{B}(x_1
  \hat{\vect{e}}_1) 
  =
  \int \frac{{\rm d}^3 k}{(2\pi)^3} ~ 
  \frac{e^{i k_1 L}-1}{i k_1} ~ \hat{\vect{e}}_1 \times 
  \tilde{\vect{B}}(\vect{k}),
\end{equation}
By using these equations, we find that the variance of $\vect{\theta}$
is given by 
\begin{equation}
\langle \vect{\theta}^2 \rangle
= 
\left( \frac{e}{m v} \right)^2 \int \frac{{\rm d}^3 k {\rm d}^3 k^{\prime}}{(2\pi)^6} ~
\frac{\left( e^{i k_1 L}-1 \right) \left( e^{i k_{1}^{\prime} L}-1 \right)}{-k_1 k_{1}^{\prime}} ~
({\delta}_{ij}-{\delta}_{i1} {\delta}_{j1}) \langle \tilde{B}_i (\vect{k}) \tilde{B}_j (\vect{k}^{\prime}) \rangle,
\label{eq:apdx1}
\end{equation}
Since the divergence of magnetic field vanishes $(k_i
\tilde{B}_i(\vect{k})=0)$ and the cosmic magnetic fields are
statistically isotropic and homogeneous, the square bracket in
eq.(\ref{eq:apdx1}) can be written as 
\begin{equation}
\langle \tilde{B}_i (\vect{k}) \tilde{B}_j (\vect{k}^{\prime}) \rangle
=
\frac{1}{2} (2\pi)^3 \delta^{(3)} \left( \vect{k}+\vect{k}^{\prime} \right)
\left[ \left( \delta_{ij} - \frac{k_i k_j}{k^2} \right) \frac{2 \pi^2}{k^3} \mathcal{P}_B (k) 
+ i \epsilon_{ijl} k_l H(k) \right],
\label{eq:apdx2}
\end{equation}
where $\mathcal{P}_B (k)$ is the magnetic power spectrum and $H(k)$
stands for the helicity component of magnetic fields~\cite{Durrer:2003ja}. 
By substituting eq.(\ref{eq:apdx2}) into (\ref{eq:apdx1}), we obtain 
\begin{eqnarray}
 \langle \vect{\theta}^2 \rangle
& = & \frac{2}{3}
\left( \frac{e L}{m v} \right)^2 \int \frac{{\rm d} k}{k} ~ 
\mathcal{P}_B (k) ~ F(kL), \label{eqn:thetatheta}\\
 F(z) 
& \equiv & \frac{3}{2}
z^{-2} \left[ \cos (z) - \frac{\sin (z)}{z} + z {\rm Si}(z) \right]
\sim \left\{
\begin{array}{l}
1 + \mathcal{O} (z^2) ~~~~~~~(z \ll 1) \\
\frac{3\pi}{4z} + \mathcal{O} (z^{-2}) ~~~(z \gg 1)
\end{array} \right. ,
\end{eqnarray}
where Si$(z)$ denotes the sine integral function. For $z \geq 0$, $F(z)$
satisfies 
\begin{equation}
 0< F(z) \leq 1, \quad 
  0 \leq zF(z) \leq \alpha, \quad 
  \alpha\equiv {\rm Max} [zF(z)] \simeq 2.48.
\end{equation}

In order to find a proper definition of the effective strength of the
magnetic field (including its normalization) for a given spectrum 
$\mathcal{P}_B(k)$, as a fiducial configuration let us consider a
homogeneous magnetic field whose direction is perpendicular to the
particle's trajectory. Denoting the strength of the fiducial magnetic
field as $B_{\perp}$, the bending angle is $\theta=L/R_{\rm L}=(eB_{\perp}L)/(mv)$, where $R_{\rm L}$ is the Larmor radius.  
On the other hand, for a statistically isotropic spectrum, the variance
of the magnetic field in three-dimensions is three halves of the
variance of the magnetic field projected onto the two-dimensional
subspace perpendicular to the particle's trajectory. Thus, it is natural
to define the effective strength of the magnetic field as 
\begin{equation}
 B_{\rm eff}^2 \equiv \frac{3}{2}\left(\frac{mv}{eL}\right)^2
  \langle \vect{\theta}^2 \rangle.
\end{equation}
Combining this with the formula (\ref{eqn:thetatheta}), we obtain 
\begin{equation}
 B_{\rm eff}^2
  = \int \frac{{\rm d} k}{k} 
  F(kL)\, \mathcal{P}_{B}(k). 
\end{equation}

\end{document}